\documentclass[prd,aps,nofootinbib,showpacs,showkeys]{revtex4}

\def\XXint#1#2#3{{\setbox0=\hbox{$#1{#2#3}{\int}$}
     \vcenter{\hbox{$#2#3$}}\kern-.5\wd0}}

\usepackage{graphicx}
\usepackage{bm}
\usepackage{amssymb}
\usepackage{amsmath}
\usepackage{euscript}

\begin{document}

\title{Quantum Fluctuations of a Coulomb Potential as a Source of
Flicker Noise}

\author{Kirill~A.~Kazakov}\email{kirill@phys.msu.ru}

\affiliation{Department of Theoretical Physics,
Physics Faculty,\\
Moscow State University, $119899$, Moscow, Russian Federation}

\begin{abstract}
The power spectrum of quantum fluctuations of the electromagnetic
field produced by an elementary particle is determined. It is
found that in a wide range of practically important frequencies
the power spectrum of fluctuations exhibits an inverse frequency
dependence. The magnitude of fluctuations produced by a conducting
sample is shown to have a Gaussian distribution around its mean
value, and its dependence on the sample geometry is determined. In
particular, it is demonstrated that for geometrically similar
samples the power spectrum is inversely proportional to the sample
volume. It is argued also that the magnitude of fluctuations
induced by external electric field is proportional to the field
strength squared. A comparison with experimental data on flicker
noise measurements in continuous metal films is made.
\end{abstract}

\maketitle

\section{Introduction}

Measurements of voltage fluctuations in various media show that at
sufficiently low frequencies, power spectra of fluctuations in all
conducting materials exhibit a universal profile which is close to
inverse frequency dependence, and called for this reason a flicker
$1/f$-noise. Although this noise is dominating only at low
frequencies, experiments show the presence of the $1/f$-component
in the whole measured band, from $10^{-6} Hz$ to $10^6 Hz.$ Also,
the following three main characteristic properties are universally
held: The $1/f$-noise produced by a conducting sample 1) is
inversely proportional to its volume, 2) is Gaussian, and 3) its
part induced by external electric field is proportional to the
field strength squared.

A number of mechanisms has been put forward to explain the origin
of the $1/f$-noise. The property 3) is typical for resistive
systems, and suggests that the flicker noise can arise from
resistance fluctuations \cite{buck}. It has been also proposed
that fluctuations in the carrier mobility \cite{hooge,klein}, or
in the number of carriers caused by surface traps \cite{mcwhorter}
might explain the origin of the $1/f$-profile of the power
spectrum. All these models, however, have restricted validity,
because they involve one or another assumption specific to the
problem under consideration. For instance, assuming that the
resistance fluctuations in the first of the above-mentioned models
spring from temperature fluctuations, one has to choose an
appropriate spatial correlation of these fluctuations in order to
obtain the $1/f$-profile of the power spectrum \cite{voss}.
Similarly, the model suggested in Ref.~\cite{mcwhorter} requires a
specific distribution of trapping times, etc. In addition to that,
the models suggested so far reproduce the $1/f$-profile only in a
restricted range of frequencies. On the other hand, the ubiquity
of the flicker noise, and universality of its properties imply
that there must exist a simple and universal, and therefore,
fundamental underlying reason. It is natural to look for this
reason in the quantum properties of charge carriers. In this
direction, the problem has been extensively investigated by Handel
and co-workers \cite{handel}. Handel's approach is based on the
theory of infrared radiative corrections in quantum
electrodynamics. Handel showed that the $1/f$ power spectrum of
photons emitted in any scattering process can be derived from the
well-know property of bremsstrahlung, namely, from the infrared
divergence of the cross-section considered as a function of the
energy loss. Thus, this theory treats the $1/f$-noise as a
relativistic effect (in fact, the noise level in this theory $\sim
\alpha (\Delta\bm{v})^2/c^2,$ where $\alpha$ is the fine structure
constant, $\Delta\bm{v}$ velocity change of the particle being
scattered, and $c$ the speed of light). It should be mentioned,
however, that the Handel's theory has been severely criticized in
many respects \cite{tremblay,kampen}.

The purpose of this paper is to draw attention to another quantum
aspect of the electromagnetic interaction, which is of a purely
nonrelativistic nature. It turns out that there is a simple and
quite general property of the electromagnetic interactions of
quantized matter, which may be the origin of flicker noise.
Namely, it will be shown below that the power spectrum of the
Coulomb field fluctuations produced by spreading wave packet of a
free charged particle exhibits an inverse frequency dependence in
the low-frequency limit. We will show also that the power spectrum
possesses the three above-mentioned characteristic properties of
flicker noise as well.

The main tool we use in investigating quantum fluctuations of the
Coulomb field produced by a quantum particle is the two-point
correlation function of the electromagnetic field. Properties of
this function in the coincidence limit were investigated in detail
in Ref.~\cite{kazakov}. It was found, in particular, that the root
mean square fluctuation of the Coulomb potential is of zeroth
order in the Planck constant $\hbar.$ It was shown, furthermore,
that the Fourier transform of the correlation function with
respect to time exhibits an inverse frequency dependence in the
low-frequency limit. One of the goals of the present paper is to
show that essentially the same formula describes also the
low-frequency limit of the power spectrum of fluctuating
electromagnetic fields measured at {\it two distinct} time
instants. The reason underlying this similarity is that in both
cases the leading contribution to the correlation function is
contained in its disconnected part.

The paper is organized as follows. In Sec.~\ref{prelim} a
preliminary consideration of the problem is given, and the
Schwinger-Keldysh formalism used throughout the work is briefly
reviewed. The low-frequency asymptotic of the power spectrum is
calculated in Sec.~\ref{calcul}. It is proved in
Sec.~\ref{connected} that the low-frequency asymptotic of the
connected part of the correlation function is logarithmic.
Contribution of the disconnected part is calculated exactly in
Sec.~\ref{discon}, and is shown to exhibit an inverse frequency
dependence. The obtained result is analyzed to verify that it is
in agreement with the above-mentioned general characteristics of
flicker noise, and then compared with experimental data in
Sec.~\ref{estim}. Section \ref{conclude} summarizes the results of
the work and states the conclusion.

\section{Preliminaries}\label{prelim}

Let us consider a single particle with mass $m$ and electric
charge $e.$ In classical theory, the electromagnetic field
produced by such particle at rest is described by the Coulomb
potential
\begin{eqnarray}\label{coulomb}&&
A_0 = \frac{e}{4\pi r}\,, \qquad \bm{A} = 0\,.
\end{eqnarray}
\noindent Our aim is to determine quantum properties of this
potential. More precisely, we will be concerned with the quantum
fluctuations of the field-theoretic counterpart of this potential
at the zeroth order in the Planck constant $\hbar.$

\begin{figure}
\includegraphics{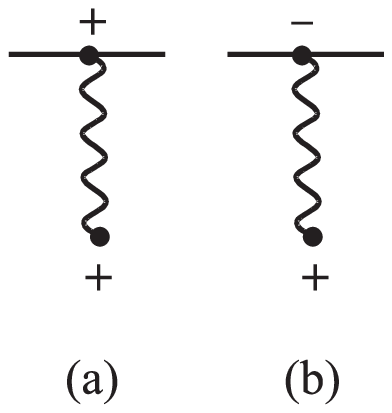}
\caption{} \label{fig1}
\end{figure}

In quantum theory, Eq.~(\ref{coulomb}) is reproduced by
calculating the corresponding mean fields, $\langle {\rm in}|
\hat{A}_0|{\rm in}\rangle,$ $\langle {\rm in}|\hat{\bm{A}}|{\rm
in} \rangle.$ In the Schwinger-Keldysh formalism
\cite{keldysh,schwinger2}, they are given by
\begin{eqnarray}\label{fint}
\langle {\rm in}| \hat{A}^{\mu} |{\rm in} \rangle = \int
\EuScript{D}\Phi_{-}\int \EuScript{D}\Phi_{+}~A^{\mu}_+\exp\{i
S[\Phi_{+}] - i S[\Phi_{-}]\}\,,
\end{eqnarray}
\noindent where the subscript $+$ ($-$) shows that the time
argument of the integration variable runs from $-\infty$ to
$+\infty$ (from $+\infty$ to $-\infty$). Integration is over all
fields satisfying
\begin{eqnarray}\label{bcond}
A^{+}_{\pm} \to 0\,, \quad \phi^{+}_{\pm} \to \phi_0^+ \,, \quad
(\phi^*)^{+}_{\pm} \to (\phi^*_0)^+  \qquad {\rm for} \quad t
&\to& -\infty\,, \nonumber\\ \Phi_{+} = \Phi_{-} \qquad {\rm for}
\quad t &\to& + \infty\,,
\end{eqnarray} \noindent
where $\Phi$ collectively denotes the fundamental fields of the
theory, $\Phi = \{A_{\mu},\phi,\phi^*\},$ the components
$\phi,\phi^*$ describing the charged particle which for simplicity
will be assumed scalar, $\phi_0$ is the given particle state,
$\EuScript{D}\Phi$ the invariant integral measure, and $S =
S[\Phi]$ the action functional of the system. Assuming that the
gradient invariance of the theory is fixed by the Lorentz
condition
\begin{eqnarray}\label{gauge}
G\equiv\partial^{\mu}A_{\mu} = 0\,,
\end{eqnarray}
\noindent the action takes the form
\begin{eqnarray}\label{action}
S[\Phi] &=& S_0[\Phi] + S_{\rm gf}[\Phi]\,, \nonumber\\
S_0[\Phi] &=& {\displaystyle\int} d^4 x
\left\{(\partial_{\mu}\phi^* + i e A_{\mu}\phi^*)
(\partial^{\mu}\phi - ie A^{\mu}\phi) - m^2 \phi^*\phi\right\} -
\frac{1}{4}{\displaystyle\int} d^4 x F_{\mu\nu} F^{\mu\nu}\,,
\nonumber\\ S_{\rm gf}[\Phi] &=& -\frac{1}{2}{\displaystyle\int}
d^4 x~G^2\,,\quad F_{\mu\nu} =
\partial_{\mu} A_{\nu} - \partial_{\nu} A_{\mu}\,.
\end{eqnarray}
\noindent Diagrammatics generated upon expanding the integral
(\ref{fint}) in powers of the coupling constant $e$ consists of
the following elements. There are four types of pairings for each
field $A_{\mu}$ or $\phi,$ corresponding to the four different
ways of placing two field operators on the two branches of the
time path. They are conveniently combined into $2 \times 2$
matrices\footnote{Below, Gothic letters are used to distinguish
quantities representing columns, matrices etc. with respect to
indices $+,-.$}
$$\mathfrak{D}^{\mu\nu}(x,y) = \left(
\begin{array}{cc}
D^{\mu\nu}_{++}(x,y)&D^{\mu\nu}_{+-}(x,y)\\
D^{\mu\nu}_{-+}(x,y)&D^{\mu\nu}_{--}(x,y)
\end{array}\right) =
\left(\begin{array}{cc}i\langle
T\hat{A}^{\mu}(x)\hat{A}^{\nu}(y)\rangle_0 & i\langle
\hat{A}^{\nu}(y)\hat{A}^{\mu}(x)\rangle_0\\
i\langle \hat{A}^{\mu}(x)\hat{A}^{\nu}(y)\rangle_0 & i\langle
\tilde{T}\hat{A}^{\mu}(x)\hat{A}^{\nu}(y)\rangle_0\end{array}\right)\,,
$$
$$\mathfrak{D}(x,y) = \left(
\begin{array}{cc}
D_{++}(x,y)&D_{+-}(x,y)\\
D_{-+}(x,y)&D_{--}(x,y)
\end{array}\right) =
\left(\begin{array}{cc}i\langle
T\hat{\phi}(x)\hat{\phi}^\dag(y)\rangle_0 & i\langle
\hat{\phi}^\dag(y)\hat{\phi}(x)\rangle_0\\i\langle
\hat{\phi}(x)\hat{\phi}^\dag(y)\rangle_0 & i\langle
\tilde{T}\hat{\phi}(x)\hat{\phi}^\dag(y)\rangle_0\end{array}\right)\,,
$$
where the operation of time ordering $T$ ($\tilde{T}$) arranges
the factors so that the time arguments decrease (increase) from
left to right, and $\langle\cdot\rangle_0$ denotes vacuum
averaging. The ``propagators'' $\mathfrak{D}_{\mu\nu},$
$\mathfrak{D}$ satisfy the following matrix equations
\begin{eqnarray}\label{aprop}
\int d^4 z~\mathfrak{G}_{\mu\nu}(x,z)\mathfrak{D}^{\nu\alpha}(z,y)
&=& - \mathfrak{e}\delta_{\mu}^{\alpha}\delta^{(4)}(x-y)\,,
\quad\mathfrak{G}_{\mu\nu}(x,y) =
\mathfrak{i}~\frac{\delta^2S^{(2)}}{\delta
A^{\mu}(x)\delta A^{\nu}(y)}\,, \\
\int d^4 z~\mathfrak{G}(x,z)\mathfrak{D}(z,y) &=& -
\mathfrak{e}\delta^{(4)}(x-y)\,, \quad\mathfrak{G}(x,y) =
\mathfrak{i}~\frac{\delta^2S^{(2)}}{\delta \phi^*(x)\delta
\phi(y)}\,,\label{aprop1}
\end{eqnarray}
\noindent where $S^{(2)}$ is the free field part of the action,
$\mathfrak{e},$ $\mathfrak{i}$ are $2\times2$ matrices with
respect to indices $+,-:$
$$\mathfrak{e} = \left(\begin{array}{cc}
1&0\\0&1\end{array}\right)\,, \quad \mathfrak{i} =
\left(\begin{array}{cc} 1&0\\0&-1\end{array}\right)\,.$$ As in the
ordinary Feynman diagrammatics of the S-matrix theory, the
propagators are contracted with the vertex factors generated by
the interaction part of the action, $S^{\rm int}[\Phi] = S[\Phi] -
S^{(2)}[\Phi],$ with subsequent summation over $(+,-)$ in the
vertices, each ``$-$'' vertex coming with an extra factor $(-1).$
This can be represented as the matrix multiplication of
$\mathfrak{D}_{\mu\nu},\mathfrak{D}$ with suitable matrix
vertices. For instance, the $A\partial\phi\phi^*$ part of the
action generates the matrix vertex $\mathfrak{V}$ which in
components has the form
$$V_{ijk}^{\mu}(x,y,z) = s_{ijk}\left.\frac{\delta^3 S}{\delta A_{\mu}(x)
\delta\phi(y)\delta\phi^*(z)}\right|_{\Phi = 0}\,,$$ where the
indices $i,j,k$ take the values $+,-,$ and $s_{ijk}$ is defined by
$s_{+++} = - s_{---} = 1$ and zero otherwise. External $\phi$
($\phi^*$) line is represented in this notation by a column (row)
$$\mathfrak{r} = \left(\begin{array}{c} \phi_0 \\ \phi_0
\end{array}\right), \quad \mathfrak{r}^\dag = (\phi^*_0,\phi^*_0),$$ satisfying
\begin{eqnarray}\label{free}
\int d^4 z~\mathfrak{G}(x,z)\mathfrak{r}(z) = \left(\begin{array}{c} 0 \\
0 \end{array}\right), \quad \int d^4 x ~\mathfrak{r}^\dag(x)
\mathfrak{G}(x,z) = (0,0)\,.
\end{eqnarray}
\noindent The tree diagrams contributing to the right hand side of
Eq.~(\ref{fint}) are depicted in Fig.~\ref{fig1}. Of these only
the diagram in Fig.~\ref{fig1}(a) gives rise to a non-zero
contribution. The momentum flow in the other diagram is
inconsistent with the momentum conservation in the vertex for any
non-zero value of the momentum transfer $p_{\mu},$ because the
lines of this diagram are all on the mass shell. The same diagram
\ref{fig1}(a) represents also the tree value of the in-out matrix
element $\langle {\rm out}|\hat{A}^{\mu}(x)|{\rm in}\rangle$
calculated using the ordinary Feynman rules. This is natural
because we are concerned here with a one-point function whose
matrix element is evaluated between one-particle states for which
$|{\rm in}\rangle$ is essentially the same as $|{\rm out}\rangle.$
The things change, however, if we want to determine the field
fluctuation. In the Schwinger-Keldysh formalism, the expectation
value of the product $\hat{A}^{\mu}(x)\hat{A}^{\nu}(x')$ has the
form
\begin{eqnarray}\label{fintctp1}
\langle {\rm in} |\hat{A}^{\mu}(x)\hat{A}^{\nu}(x')|{\rm
in}\rangle = \int \EuScript{D}\Phi_{-}\int
\EuScript{D}\Phi_{+}~A^{\mu}_-(x)A^{\nu}_+(x')\exp\{i S[\Phi_{+}]
- i S[\Phi_{-}]\}\,,
\end{eqnarray}
\noindent  The corresponding tree diagrams are shown in
Fig.~\ref{fig2}. We see that these diagrams {\it differ} from
those we would have obtained applying the ordinary Feynman rules
to the quantity $\langle {\rm out}|
\hat{A}^{\mu}(x)\hat{A}^{\nu}(x')|{\rm in}\rangle.$ We define the
correlation function of the electromagnetic field fluctuations by
\begin{eqnarray}\label{corr}&&
C_{\mu\nu}(x,x') = \langle {\rm
in}|\hat{A}_{\mu}(x)\hat{A}_{\nu}(x')|{\rm in}\rangle - \langle
{\rm in}|\hat{A}_{\mu}(x)|{\rm in}\rangle\langle {\rm
in}|\hat{A}_{\nu}(x')|{\rm in}\rangle.
\end{eqnarray}
\noindent The first term in this definition is usually written in
a symmetrical form thus rendering the function $C_{\mu\nu}(x,x')$
real. However, this makes no difference for what follows because
it will be shown below that the leading low-frequency term of the
correlation function is contained entirely in its disconnected
part [the second term in Eq.~(\ref{corr})].

We are concerned with the power spectrum of correlations in the
values of the electromagnetic fields measured at two distinct time
instants (spatial separation between the observation points,
$|\bm{x} - \bm{x}'|,$ is also kept arbitrary). Thus, fixing one of
the time arguments, say, $t',$ we define the Fourier transform of
$C_{\mu\nu}(x,x')$ with respect to $(t - t')$
\begin{eqnarray}\label{corrf}&&
C_{\mu\nu}(\bm{x},\bm{x}',t',\omega) =
\int\limits_{-\infty}^{+\infty}dt C_{\mu\nu}(x,x')e^{i\omega(t -
t')}\,.
\end{eqnarray}
\noindent To complete the present section, we write out explicit
expressions for various pairings of the photon and scalar fields
\begin{eqnarray}\label{explprop}
\mathfrak{D}_{\mu\nu} &=& - \eta_{\mu\nu}\mathfrak{D}^{0}\,,
\qquad
\mathfrak{D}^{0} \equiv \mathfrak{D}|_{m=0}\,, \nonumber\\
D_{++}(x,y) &=& \int\frac{d^4 k}{(2\pi)^4}\frac{e^{-ik(x-y)}}{m^2
- k^2 - i0}\,, \quad D_{--}(x,y) =
\int\frac{d^4 k}{(2\pi)^4}\frac{e^{-ik(x-y)}}{k^2 - m^2 - i0}\,,\nonumber\\
D_{-+}(x,y) &=& i\int\frac{d^4 k}{(2\pi)^3}\theta(k^0)\delta(k^2 -
m^2)e^{-ik(x-y)}\,, \quad D_{+-}(x,y) =  D_{-+}(y,x)\,.
\end{eqnarray}
\noindent

\section{Evaluation of the leading contribution}\label{calcul}

Evaluation of the low-frequency asymptotic of the correlation
function proceeds in two steps. First, we will prove in
Sec.~\ref{connected} that the low-frequency asymptotic of the
connected part of $C_{\mu\nu}$ [the first term in
Eq.~(\ref{corr})] is only logarithmic. The contribution of the
disconnected part will be calculated exactly in Sec.~\ref{discon}.
It will be shown that this contribution exhibits an inverse
frequency dependence, and thus dominates in the low-frequency
limit.

\subsection{Low-frequency asymptotic of the connected part of
correlation function}\label{connected}

\begin{figure}
\includegraphics{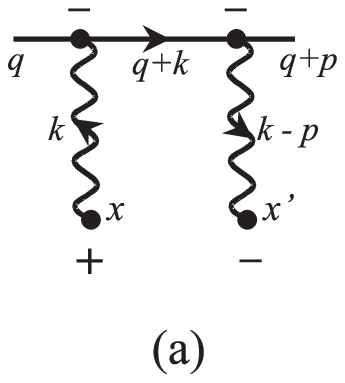}
\includegraphics{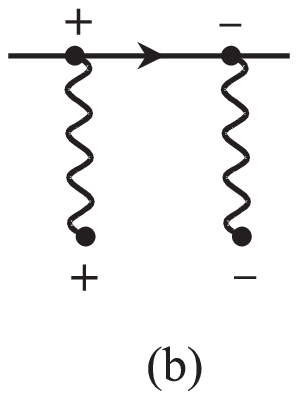}
\includegraphics{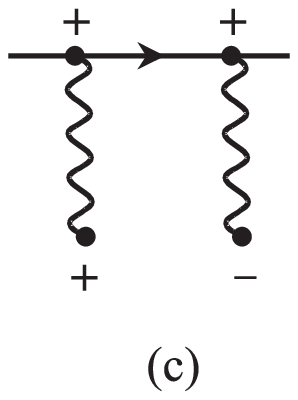}
\includegraphics{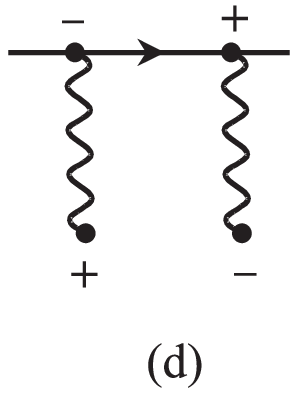}
\includegraphics{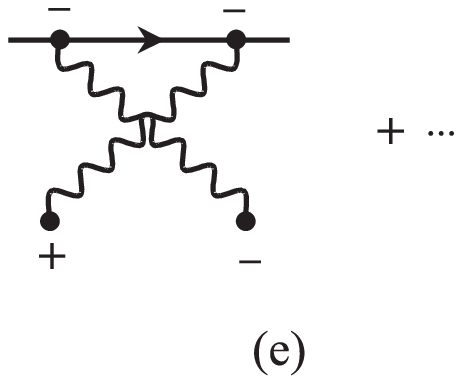}
\includegraphics{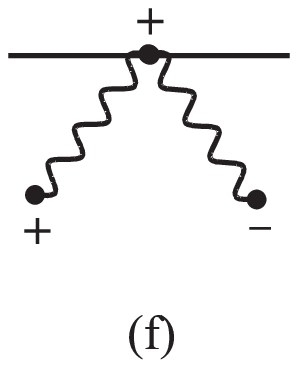}
\includegraphics{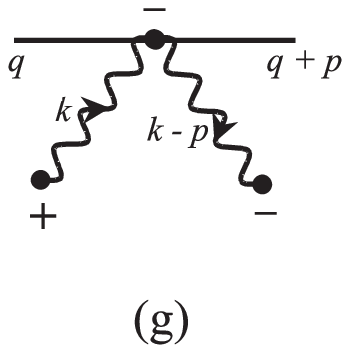}
\caption{} \label{fig2}
\end{figure}

Before going into detailed calculations, let us first exclude the
diagrams in Fig.~\ref{fig2} which do not contain the $\hbar^0$
contribution. It is not difficult to see that such are the
diagrams without internal matter line. Indeed, consider, for
instance, the diagram (g). It is proportional to the integral
$$\int d^4 k
\theta(k^0)\delta(k^2)\frac{e^{ik(x-x')}}{(k-p)^2}\ ,$$ which does
not involve the particle mass at all. Taking into account that
each external scalar line gives rise to the factor
$(2\varepsilon_{\bm{q}})^{-1/2},$ where $\varepsilon_{\bm{q}} =
\sqrt{m^2 + \bm{q}^2}\approx m\,,$ we see that the contribution of
the diagram \ref{fig2}(g) is proportional to $1/m.$ Hence, on
dimensional grounds, this diagram is proportional to $\hbar.$ The
same is true of all other diagrams without internal matter lines.

The sum of diagrams (a)--(e) in Fig.~\ref{fig2} has the following
symbolic form
$$I_{\mu\nu} + I^{\rm tr}_{\mu\nu}\,, \qquad I_{\mu\nu} =
\frac{1}{i}\left\{\mathfrak{D}_{\mu\alpha}\left[\mathfrak{r}^\dag
\mathfrak{V}^{\alpha}\mathfrak{D}\mathfrak{V}^{\beta}\mathfrak{r}\right]
\mathfrak{D}_{\beta\nu}\right\}_{+-}\,,$$ where the superscript
``tr'' means transposition of the indices and spacetime arguments
referring to the points of observation: $\mu\leftrightarrow\nu,$
$+\leftrightarrow -,$ $x\leftrightarrow x'$ [the transposed
contribution is represented by the diagrams collected in part (e)
of Fig.~\ref{fig2}]. Written longhand, $I_{\mu\nu}$ reads
\begin{eqnarray}\label{diagen}
I_{\mu\nu}(x,x') = ie^2\iint d^4 z d^4 z'\Biggl\{&+&
D^0_{++}(x,z)\left[\phi_0^*(z)
\stackrel{\leftrightarrow}{\partial_{\mu}}
D_{++}(z,z')\stackrel{\leftrightarrow}{\partial_{\nu}^{\,\prime}}
\phi_0(z')\right]D^0_{+-}(z',x') \nonumber\\ &-&
D^0_{++}(x,z)\left[\phi_0^*(z)
\stackrel{\leftrightarrow}{\partial_{\mu}}
D_{+-}(z,z')\stackrel{\leftrightarrow}{\partial_{\nu}^{\,\prime}}
\phi_0(z')\right]D^0_{--}(z',x')\nonumber\\ &-&
D^0_{+-}(x,z)\left[\phi_0^*(z)
\stackrel{\leftrightarrow}{\partial_{\mu}}
D_{-+}(z,z')\stackrel{\leftrightarrow}{\partial_{\nu}^{\,\prime}}
\phi_0(z')\right]D^0_{+-}(z',x')\nonumber\\ &+&
D^0_{+-}(x,z)\left[\phi_0^*(z)
\stackrel{\leftrightarrow}{\partial_{\mu}}
D_{--}(z,z')\stackrel{\leftrightarrow}{\partial_{\nu}^{\,\prime}}
\phi_0(z')\right]D^0_{--}(z',x') \Biggr\}\,,\nonumber\\
\end{eqnarray}
\noindent where
$$\varphi\stackrel{\leftrightarrow}{\partial_{\mu}}\psi =
\varphi\partial_{\mu}\psi - \psi\partial_{\mu}\varphi\,.$$
Contribution of the third term in the right hand side of
Eq.~(\ref{diagen}) is zero identically. Indeed, using
Eq.~(\ref{explprop}), and performing spacetime integrations we see
that the three lines coming, say, into $z$-vertex are all on the
mass shell, which is inconsistent with the momentum conservation
in the vertex. Taking the Fourier transform of $I_{\mu\nu}(x,x'),$
$$\tilde{I}_{\mu\nu}(\bm{x},\bm{x}',t',\omega) =
\int\limits_{-\infty}^{+\infty}d t
I_{\mu\nu}(x,x')e^{i\omega(t-t')}\,,$$ the remaining terms in
Eq.~(\ref{diagen}) take the form
\begin{eqnarray}\label{diagenk1}
\tilde{I}_{\mu\nu}(\bm{x},\bm{x}',t',\omega) &=& e^2\iint
\frac{d^3 \bm{q}}{(2\pi)^3} \frac{d^3
\bm{p}}{(2\pi)^3}\frac{a^*(\bm{q})a(\bm{q} +
\bm{p})}{\sqrt{2\varepsilon_{\bm q}2\varepsilon_{{\bm q} +
\bm{p}}}} e^{-ip^0(t' - t_0) + i\bm{p}\bm{x}'}
\tilde{J}_{\mu\nu}(p,q,\bm{x}-\bm{x}',\omega)\,, \\
p^0 &=& \varepsilon_{\bm{q} + \bm{p}} - \varepsilon_{\bm{q}}\,,
\nonumber
\end{eqnarray}
\noindent where
\begin{eqnarray}\label{diagenk3}&&
\tilde{J}_{\mu\nu}(p,q,\bm{x}-\bm{x}',\omega) = - i\int \frac{d^3
\bm{k}}{(2\pi)^3} e^{i\bm{k}(\bm{x}-\bm{x}')}(2q_{\mu} +
k_{\mu})(2q_{\nu} + k_{\nu} +
p_{\nu})\nonumber\\&&\times\Bigl\{D^0_{++}(k)D_{++}(q+k)D^0_{+-}(k-p)
-D^0_{++}(k)D_{+-}(q+k)D^0_{--}(k-p) \nonumber\\&& +
D^0_{+-}(k)D_{--}(q+k)D^0_{--}(k-p)\Bigr\}_{k^0=\omega}\,.
\end{eqnarray}
\noindent Here $\varepsilon_{\bm{q}} = \sqrt{\bm{q}^2 + m^2}$ is
the particle's energy, $m,$ $e,$ $q_{\mu},$ and $a(\bm{q})$ are
its mass, electric charge, 4-momentum, and momentum wave function
at some time instant $t_0,$ $p_{\mu}$ 4-momentum transfer. The
function $a(\bm{q})$ is normalized by
\begin{eqnarray}\label{norm}
\int\frac{d^3 \bm{q}}{(2\pi)^3}|a(\bm{q})|^2 = 1\,,
\end{eqnarray}\noindent and is generally of the form
\begin{eqnarray}\label{abrel}
a(\bm{q}) = b(\bm{q}) e^{-i\bm{q}\bm{x}_0}\,,
\end{eqnarray}\noindent
where $\bm{x}_0$ is the particle mean position, and $b(\bm{q})$
describes the momentum space profile of the particle wave packet.
For simplicity, the particle charge distribution will be assumed
spherically-symmetric in what follows. Then $b(\bm{q})$ is a
function of $\bm{q}^2:$ $b(\bm{q}) = \beta (\bm{q}^2).$

As in Ref.~\cite{kazakov}, we work within the long-range expansion
which implies, in particular, that the spatial separations between
the field-producing particle (more precisely, its mean position)
and the points of observation, $r = |\bm{x} - \bm{x}_0|,$ $r' =
|\bm{x}' - \bm{x}_0|,$ are such that $r \bar{q}\gg 1,$
$r'\bar{q}\gg 1,$ where $\bar{q} = \sqrt{\langle\bm{q}^2\rangle}$
is the particle momentum variance (note that $\bar{q}$ is
time-independent for a free particle). To the leading order of the
long-range expansion, $p_{\mu},$ $k_{\mu}$ in the vertex factors
in Eq.~(\ref{diagenk3}) can be neglected in comparison with
$q_{\mu}.$

Let us now show that the low-frequency asymptotic of
$\tilde{J}_{\mu\nu}$ is weaker than $1/\omega.$ The first and the
third terms in the integrand in Eq.~(\ref{diagenk3}) give rise to
a contribution which is finite at $\omega = 0.$ We have for the
third term
\begin{eqnarray}&&
\int d^3 \bm{k}\left.e^{i\bm{k}(\bm{x}-\bm{x}')}
D^0_{+-}(k)D_{--}(q+k)D^0_{--}(k-p)\right|_{k^0=0} \nonumber\\&& =
\int d^2 o\int\limits_{0}^{\infty} d|\bm{k}| \frac{\bm{k}^2
e^{i\bm{k}(\bm{x}-\bm{x}')} 2\pi i \delta(\bm{k}^2)}{[-2(\bm{qk})
- i0] [p^2 + 2(\bm{pk}) - i0]} = - \frac{\pi i}{4 |\bm{q}|
p^2}\int \frac{d\phi d\theta \sin\theta}{\cos\theta + i0} = -
\frac{\pi^3}{2 |\bm{q}| p^2} \,. \nonumber
\end{eqnarray}
\noindent Similarly,
\begin{eqnarray}&&
\int d^3 \bm{k}\left.e^{i\bm{k}(\bm{x}-\bm{x}')}
D^0_{++}(k)D_{++}(q+k)D^0_{+-}(k-p)\right|_{k^0=0} \nonumber\\&& =
\int d^2 o\int\limits_{0}^{\infty} d|\bm{k}| \frac{\bm{k}^2
e^{i\bm{k}(\bm{x}-\bm{x}')}2\pi i \theta(p^0)\delta\left((p^0)^2 -
(\bm{k} - \bm{p})^2\right)}{[\bm{k}^2 - i0] [\bm{k}^2 + 2(\bm{qk})
- i0]} \nonumber
\end{eqnarray}
\noindent is also finite for all $\bm{p}$ and $\bm{q}\ne 0$ [the
singularity at $\bm{q} = 0$ or $\bm{p} = 0$ in these expressions
is inessential as it is removed by the factors $\bm{q}^2$ and
$\bm{p}^2$ in the integral measure in Eq.~(\ref{diagenk1})]. The
second term in the integrand, however, leads to a divergence at
$\omega =0.$ This divergence comes from integration over small
$|\bm{k}|.$ Hence, it is the same as the divergence of the
integral
\begin{eqnarray}
\int d^2 o\int\limits_{0}^{\infty} d|\bm{k}|
\frac{\bm{k}^2\delta(\omega^2 + 2m\omega - \bm{k}^2 - 2(\bm{qk})
)}{[-\omega^2 + \bm{k}^2 - i0][p^2 - i0]} = \frac{\pi}{|\bm{q}|
p^2 }\int\limits_{\sqrt{\bm{q}^2 + \omega^2 + 2m\omega} -
|\bm{q}|}^{\sqrt{\bm{q}^2 + \omega^2 + 2m\omega} + |\bm{q}|}
\frac{u d u }{-\omega^2 + u^2 - i0}\,. \nonumber
\end{eqnarray} \noindent The latter integral diverges for $\omega \to 0$
only logarithmically.

\subsection{Low-frequency asymptotic of the disconnected part of
correlation function}\label{discon}

Let us turn to the disconnected part of the correlation function.
To find its Fourier transform, we have to evaluate the following
integral
\begin{eqnarray}\label{meanfourier}
\int\limits_{-\infty}^{+\infty}dt\, \langle {\rm
in}|A_{\mu}(x)|{\rm in}\rangle e^{i\omega(t-t')} =
q_{\mu}e^{-i\omega t'}\tilde{I}(\bm{r},\omega)\,,
\end{eqnarray}\noindent
where
\begin{eqnarray}
\tilde{I}(\bm{r},\omega) &=& \int\limits_{-\infty}^{+\infty}dt
\left\{ e^{i\omega t}
\iint\frac{d^3\bm{q}}{(2\pi)^3}\frac{d^3\bm{p}}{(2\pi)^3}
\frac{e^{i\bm{p}\bm{r}}}{\bm{p}^2}e^{-ip^0(t-t_0)}b^*(\bm{q})
b(\bm{q} + \bm{p})\right\}\,, \quad \bm{r} = \bm{x} -
\bm{x}_0\,.\nonumber
\end{eqnarray}\noindent
Substituting $$p^0 \approx \frac{(\bm{p} + \bm{q})^2}{2m} -
\frac{\bm{q}^2}{2m}$$ gives
\begin{eqnarray}
\tilde{I}(\bm{r},\omega) = 2\pi e^{i\omega t_0}
\iint\frac{d^3\bm{q}}{(2\pi)^3}\frac{d^3\bm{p}}{(2\pi)^3}
\delta\left(\omega - \frac{\bm{p}^2 + 2 \bm{pq}}{2m}\right)
\frac{e^{i\bm{p}\bm{r}}}{\bm{p}^2}b^*(\bm{q})b(\bm{q} +
\bm{p})\,.\nonumber
\end{eqnarray}\noindent
For a spherically-symmetric charge distribution, $b(\bm{q}) =
\beta(\bm{q}^2),$ $\tilde{I}$ is actually a function of $r =
|\bm{r}|,$ $\tilde{I}(\bm{r},\omega) = \tilde{I}(r,\omega),$ and
hence, averaging over directions of $\bm{r},$ one can write
\begin{eqnarray}\label{fourier3}
\tilde{I}(r,\omega) &=& 2\pi e^{i\omega t_0}
\iint\frac{d^3\bm{q}}{(2\pi)^3}\frac{d^3\bm{p}}{(2\pi)^3}
\delta\left(\omega - \frac{\bm{p}^2 + 2 \bm{pq}}{2m}\right)
\frac{\sin(pr)}{p^3 r}\beta^*(\bm{q}^2)\beta(\bm{q}^2 + 2m\omega)
\nonumber\\ &=& \frac{me^{i\omega t_0}}{2\pi r}
\int\frac{d^3\bm{q}}{(2\pi)^3}\beta^*(\bm{q}^2)\beta(\bm{q}^2 +
2m\omega)\int\limits_{0}^{\pi}d\theta \frac{\sin\theta\sin(u
r)}{u\sqrt{\bm{q}^2 \cos^2\theta + 2m\omega}}\,, \nonumber\\ u &=&
- |\bm{q}|\cos\theta + \sqrt{\bm{q}^2 \cos^2\theta + 2m\omega}\,,
\end{eqnarray}\noindent
where $\theta$ is the angle between the vectors $\bm{p},
\bm{q}\,,$ and it is assumed that $\omega >0.$ Taking $u$ as the
integration variable yields
\begin{eqnarray}\label{fourier4}
\tilde{I}(r,\omega) &=& \frac{me^{i\omega t_0}}{8\pi^3 r}
\int\limits_{0}^{+\infty}d\bm{q}^2 \beta^*(\bm{q}^2)\beta(\bm{q}^2
+ 2m\omega)\int\limits_{\sqrt{\bm{q}^2 + 2m\omega} - |\bm{q}|
}^{\sqrt{\bm{q}^2 + 2m\omega} + |\bm{q}|}d u \frac{\sin(u
r)}{u^2}\,.
\end{eqnarray}\noindent
To further transform this integral, it is convenient to define a
function $\Gamma(\bm{q}^2,\omega)$ according to
\begin{eqnarray}\label{fourier5}
\Gamma(\bm{q}^2,\omega) =
\frac{1}{4\pi^2}\int\limits_{\bm{q}^2}^{+\infty} dz
\beta^*(z)\beta(z + 2m\omega)\,.
\end{eqnarray}\noindent
Then integrating by parts, and taking into account that
$\Gamma(\bm{q}^2,\omega)\to 0$ for $\bm{q}^2 \to \infty,$ brings
Eq.~(\ref{fourier4}) to the form
\begin{eqnarray}\label{fourier6}
\tilde{I}(r,\omega) &=& \frac{me^{i\omega t_0}}{2\pi r}
\int\limits_{0}^{+\infty}d|\bm{q}|
\frac{\Gamma(\bm{q}^2,\omega)}{\sqrt{\bm{q}^2 + 2m\omega
}}\biggl\{\left.\frac{\sin(u r)}{u}\right|_{u = \sqrt{\bm{q}^2 +
2m\omega} + |\bm{q}|} \nonumber\\&& + \left.\frac{\sin(u
r)}{u}\right|_{u = \sqrt{\bm{q}^2 + 2m\omega} -
|\bm{q}|}\biggr\}\,.
\end{eqnarray}\noindent
Finally, integrating $\tilde{I}$ by parts once more, we find
\begin{eqnarray}
\tilde{I}(r,\omega) &=& \frac{e^{i\omega t_0}}{2\pi r\omega}
\int\limits_{0}^{+\infty}d|\bm{q}|~\sin\left(r\sqrt{\bm{q}^2 +
2m\omega }\right)\left\{2\Gamma\cos(|\bm{q}|r) +
\frac{1}{r}\frac{\partial\Gamma}{\partial |\bm{q}|}
\sin(|\bm{q}|r) \right\}\,. \label{i2}
\end{eqnarray}\noindent
This exact expression considerably simplifies in the practically
important case of low $\omega$ and large $r.$ Namely, if $\omega$
is such that
\begin{eqnarray}
\omega\ll \frac{\bar{q}}{m r} \equiv \omega_0\,, \label{cond}
\end{eqnarray}\noindent and also $$r\bar{q} \gg 1$$
(and therefore, $\omega \ll \bar{q}^2/m$), then the first term in
$\tilde{I}$ turns out to be exponentially small ($\sim e^{-r
\bar{q}}$) because of the oscillating product of trigonometric
functions. Replacing $\sin^2(qr)$ by its average value (1/2) in
the rest of $\tilde{I}$ gives
\begin{eqnarray}
\tilde{I}(r,\omega) &=& \frac{e^{i\omega t_0}}{4\pi r^2\omega}
\int\limits_{0}^{+\infty}d|\bm{q}|\frac{\partial\Gamma}{\partial
|\bm{q}|} = - \frac{e^{i\omega t_0}\Gamma(0,0)}{4\pi r^2\omega} =
- \frac{e^{i\omega t_0}}{16\pi^3
r^2\omega}\int\limits_{0}^{+\infty} dz |\beta(z)|^2 \,.
\label{i22}
\end{eqnarray}\noindent
Substituting the obtained expression into the defining equations
(\ref{meanfourier}), (\ref{corr}), (\ref{corrf}), we thus obtain
the following expression for the low-frequency asymptotic of the
correlation function
\begin{eqnarray}\label{main}
C_{00}(\bm{x},\bm{x}',t',\omega) &=& e^{i\omega (t_0 - t')
}\frac{e^2}{16\pi^3 r^2\omega}\int\limits_{0}^{+\infty} dz
|\beta(z)|^2 \iint \frac{d^3 \bm{q}}{(2\pi)^3} \frac{d^3
\bm{p}}{(2\pi)^3}\frac{e^{-ip(x' - x_0)}}{\bm{p}^2}
\beta^*(\bm{q}^2)\beta((\bm{q} + \bm{p})^2)\,, \nonumber\\
\end{eqnarray}\noindent all other components of $C_{\mu\nu}$
being suppressed by the factor $|\bm{q}|/m\ll 1.$ For the time
instants $t'$ such that $r, r' \gg D_{t'},$ where $D_{t'}$ is a
characteristic linear dimension of the particle charge
distribution (for instance, variance of the particle coordinates),
Eq.~(\ref{main}) simplifies to
\begin{eqnarray}\label{main1}
C_{00}(\bm{x},\bm{x}',t',\omega) &=& e^{i\omega (t_0 - t')
}\frac{e^2}{64\pi^4 r^2 r' \omega}\int\limits_{0}^{+\infty} dz
|\beta(z)|^2 \,.
\end{eqnarray}\noindent

In applications to microelectronics, $\omega$ varies from $10^{-6}
{\rm Hz}$ to $10^6\,{\rm Hz},$ the relevant distances $r$ are
usually $10^{-5}\,{\rm cm}$ to $10^{-2}\,{\rm cm},$ $\bar{q}\sim
\hbar/d,$ where $d\approx 10^{-8}\,{\rm cm}$ is the lattice
spacing, and $m$ is the effective electron mass, $m\approx
10^{-27}\,{\rm g},$ hence, $\omega_0 \approx 10^{10}\,{\rm Hz},$
so the conditions $r\gg d,$ $\omega\ll\omega_0$ are always
well-satisfied.

Equation~(\ref{main1}) represents an individual contribution of a
conduction electron to the electric potential fluctuation.
Considering a large number of uncorrelated electrons in a sample,
one should take into account that the corresponding time instants
$t_0$ are distributed uniformly. Because of the oscillating
exponent $e^{i\omega (t_0 - t')},$ the magnitude of the total
noise remains at the level of the individual contribution
independently of the number of electrons. Therefore, summing up
all contributions amounts simply to averaging over $\bm{x}_0:$
\begin{eqnarray}\label{main2}
\left|C_{00}^{\rm tot}(\bm{x},\bm{x}',\omega)\right| =
\frac{e^2}{64\pi^4 |\omega|\Omega}\int\limits_{0}^{+\infty} dz
|\beta(z)|^2\int\limits_\Omega \frac{d^3\bm{x}_0}{r^2 r'}\,,
\end{eqnarray}\noindent where $\Omega$ is the sample volume. To
visualize it, one can say that various uncompensated individual
contributions ``flicker'' in various points of the sample.

It is clear from the above discussion that the distribution of the
noise magnitude around the value $C_{00}^{\rm tot}$ is Gaussian,
by virtue of the central limiting theorem. Thus, the quantum field
fluctuations in a sample possess the above-mentioned property 2).

Next, let us consider the case when the sample is in an external
electric field, $\bm{E}.$ Under the influence of the external
field, both the electron wave functions and the statistical
probability distribution are changed. It turns out, however, that
in practice, the change of the wave functions is negligible.
Namely, a direct calculation shows that the relative value of the
correction is of the order
$$\kappa_q = \frac{\hbar^2 m e|\bm{E}|}{\bar{q}^4 L}\,,$$
where $L$ is the smallest linear dimension of the sample. Using
the above estimates for $\bar{q}, L$ in the condition $\kappa_q
\ll 1$ gives $|\bm{E}| \ll 10^{10},$ in the $cgs$ system of units.
In applications to microelectronics, the field strength is usually
$|\bm{E}| \lesssim 1,$ so $\kappa_q \approx 10^{-10}.$ On the
other hand, the relative change of the statistical distribution
function, given by the kinetic theory, is of the order
$$\kappa_s = \frac{e|\bm{E}|\,l}{\bar{\varepsilon}}\,,$$
where $l$ is the electron mean free path, and $\bar{\varepsilon}$
its mean energy, $\bar{\varepsilon}\approx \hbar^2/m d^2.$
$\kappa_s$ is generally not small. Furthermore, the statistical
distribution is generally analytic in $\bm{E}$ in a vicinity of
$\bm{E} = 0,$ and hence the leading correction to the scalar
quantity $C_{00}$ is proportional to $\bm{E}^2$ [property 3)].

Because of collisions with phonons and impurities in the crystal,
evolution of the conduction electrons usually cannot be considered
free. It is important, however, that in view of the smallness of
the electron mass in comparison with the mass of the lattice
atoms, these collisions may often be considered elastic. In such
collisions, the electron energy is not changed, and therefore, so
is time evolution of the electron wave function. Hence, the above
results concerning dispersion of the electromagnetic field
fluctuations remain essentially the same upon account of the
electron collisions, except that now they must be expressed in
terms of the electron density matrix, rather than wave function.
Denoting the momentum space density matrix of the electron by
$\rho(\bm{p}^2,\bm{q}^2),$ Eq.~(\ref{main2}) thus takes the form
\begin{eqnarray}\label{main3}
\left|C_{00}^{\rm tot}(\bm{x},\bm{x}',\omega)\right| =
\frac{e^2\Gamma_0}{16\pi^2 |\omega|\Omega}\int\limits_\Omega
\frac{d^3\bm{x}_0}{r^2 r'}\,, \quad \Gamma_0 \equiv
\frac{1}{4\pi^2}\int\limits_{0}^{+\infty} dz \rho(z,z)\,.
\end{eqnarray}\noindent The quantity denoted here by $\Gamma_0$ is
nothing but the expectation value of the inverse particle
momentum: $$\Gamma_0 = \int\frac{d^3\bm{q}}{(2\pi)^3}
\frac{\rho(\bm{q}^2,\bm{q}^2)}{|\bm{q}|} = \left\langle
\frac{1}{|\bm{q}|}\right\rangle\,.$$ Hence, by the order of
magnitude, $\Gamma_0 \approx 1/\bar{q}$ (in the ordinary units,
$\Gamma_0 \approx \hbar/\bar{q}.$)

In practice, one is interested in fluctuations of the voltage,
$U,$ between two leads attached to a sample. Using the above
results, it is not difficult to write down an expression for the
voltage correlation function, $C_{\rm U}(\bm{x},\bm{x}',\omega)$.
We have
\begin{eqnarray}&&\label{comment}
\langle {\rm in}|U(\bm{x},\bm{x}',t)|{\rm in}\rangle \langle {\rm
in}|U(\bm{x},\bm{x}',t')|{\rm in}\rangle = \langle {\rm
in}|A_0(\bm{x},t) - A_0(\bm{x}',t)|{\rm in}\rangle \langle {\rm
in}|A_0(\bm{x},t') - A_0(\bm{x}',t')|{\rm in}\rangle \nonumber\\&&
= \langle {\rm in}|A_0(\bm{x},t)|{\rm in}\rangle \langle {\rm
in}|A_0(\bm{x},t')|{\rm in}\rangle + \langle {\rm
in}|A_0(\bm{x}',t)|{\rm in}\rangle \langle {\rm
in}|A_0(\bm{x}',t')|{\rm in}\rangle \nonumber\\ && - \langle {\rm
in}|A_0(\bm{x}',t)|{\rm in}\rangle \langle {\rm
in}|A_0(\bm{x},t')|{\rm in}\rangle - \langle {\rm
in}|A_0(\bm{x},t)|{\rm in}\rangle \langle {\rm
in}|A_0(\bm{x}',t')|{\rm in}\rangle\,.
\end{eqnarray}\noindent Fourier transforming and applying
Eq.~(\ref{main1}) to each of the four terms in this expression, we
obtain the low-frequency asymptotic of the overall voltage
fluctuation across the sample
\begin{eqnarray}\label{main4}
\left|C_{\rm U}^{\rm tot}(\bm{x},\bm{x}',\omega)\right| =
\frac{e^2\Gamma_0 G}{16\pi^2 |\omega|\Omega}\,,
\end{eqnarray}\noindent where
\begin{eqnarray}\label{gfactor}
G \equiv \int\limits_\Omega d^3\bm{x}_0\left(\frac{1}{r^3} +
\frac{1}{r'^3} - \frac{1}{r^2 r'} - \frac{1}{r'^2 r} \right)\,.
\end{eqnarray}\noindent

\begin{figure}
\includegraphics{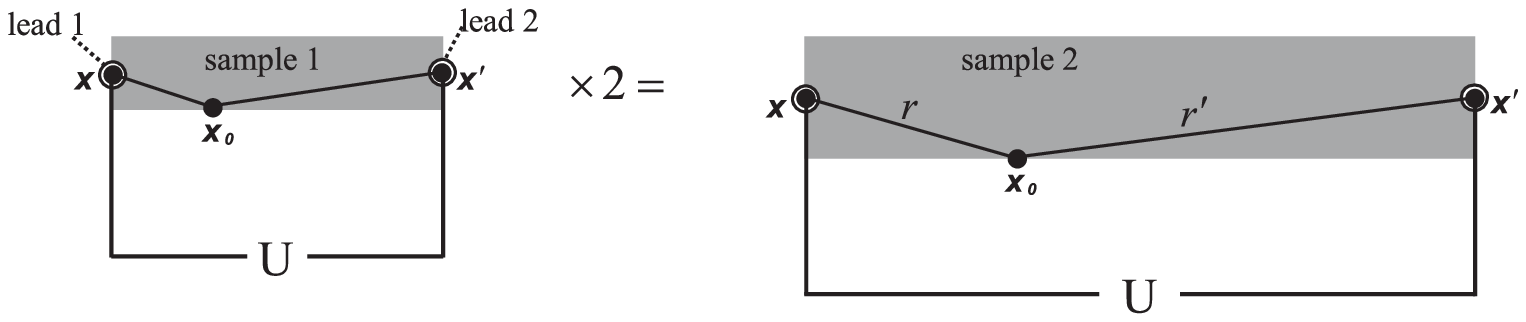}
\caption{} \label{fig3}
\end{figure}

Let us discuss the role of the sample geometry in somewhat more
detail. Consider two geometrically similar samples, and let $n$ be
the ratio of their linear dimensions (see Fig.~\ref{fig3}). Such
samples are characterized by the same value of the $G$-factor.
Indeed, the voltage across each sample is measured via two leads
attached to its surface. Hence, the radius-vectors of the leads
drawn from the center of similitude scale by the same factor $n,$
and therefore,
\begin{eqnarray}
\int\limits_{\Omega_2} \frac{d^3\bm{x}_0}{r^2 r'} &=&
\int\limits_{n^3\Omega_1} \frac{d^3\bm{x}_0}{|n\bm{x} -
\bm{x}_0|^2 |n\bm{x}' - \bm{x}_0|} = \int\limits_{\Omega_1}
\frac{n^3d^3(\bm{x}_0)}{|n\bm{x} - n\bm{x}_0|^2 |n\bm{x}' -
n\bm{x}_0|} \nonumber\\ &=& \int\limits_{\Omega_1}
\frac{d^3\bm{x}_0}{|\bm{x} - \bm{x}_0|^2 |\bm{x}' - \bm{x}_0|} =
\int\limits_{\Omega_1} \frac{d^3\bm{x}_0}{r^2 r'} \,,\nonumber
\end{eqnarray}\noindent
and likewise for the other terms in Eq.~(\ref{gfactor}). Thus, it
follows from Eq.~(\ref{main2}) that for geometrically similar
samples, the noise level is inversely proportional to the sample
volume [property 1)].

It is worth also to make the following comment concerning
Eq.~(\ref{gfactor}). The integrand in this formula involves the
terms $1/r^3$ and $1/r'^3$ which give rise formally to a
logarithmic divergence when the observation points approach the
sample. In this connection, it should be recalled that the above
calculations have been carried out under the condition
$r\bar{q}\gg 1,$ hence, $r,r'$ cannot be taken too small.
Furthermore, one should remember that in any field measurement in
a given point, one deals actually with the field averaged over a
small but finite domain surrounding this point, i.e., the voltage
lead in our case. Thus, for instance, the quantity $A_0(\bm{x},t)$
appearing in the expression (\ref{comment}) is to be substituted
by $$\EuScript{A}_0 = \frac{1}{\Omega_{\rm
L}}\int\limits_{\Omega_{\rm L}}d^3 \bm{x}A_0(\bm{x},t)\,,$$ where
$\Omega_{\rm L}$ is the voltage lead volume. As the result of this
substitution, the term $1/r^3$ takes the form
$$\frac{1}{\Omega^2_{\rm L}}\iint\limits_{\Omega_{\rm L}}\frac{d^3
\bm{x} d^3\tilde{\bm{x}}}{|\bm{x} - \bm{x}_0|^2 |\tilde{\bm{x}} -
\bm{x}_0|} \,.$$ This is finite however the lead is attached to
the sample. Equation~(\ref{gfactor}) is recovered in the case when
$r,r'$ are much larger than the sample size.

\subsection{A preliminary comparison with experiment}\label{estim}

For a preliminary comparison of the noise level given by
Eq.~(\ref{main4}) with experimental data, let us use the results
of the classic paper by Voss and Clarke \cite{voss} where the
$1/f$-noise in continuous Bi and Au films was measured. The
bismuth and gold samples had dimensions $1000\,$\AA $\times
10\,{\rm \mu m} \times 120\,{\rm \mu m}$ and $250\,$\AA $\times
8\,{\rm \mu m} \times 625\,{\rm \mu m},$ respectively. The
spectrum power for $f = \omega/2\pi = 1Hz$ and $|\bm{E}| = 0$ was
found to be about $10^{-14} V^2/Hz$ in the case of Bi sample, and
about $10^{-16} V^2/Hz$ in the case of Au sample (it is assumed
here that the contributions from external pickup and amplifier
noise in the experimental setup of Ref.~\cite{voss} do not
dominate over the sample noise component).

For a rough estimate of the right hand side of Eq.~(\ref{main4})
we take $\Gamma_0 \approx \hbar/\bar{q} \approx d \approx
10^{-8}\,{\rm cm},$ and note that $\Omega \approx 10^{-10}\,{\rm
cm^3}$ for both samples. However, the $G$-factor takes different
values for the two sample configurations. It is seen from
Eq.~(\ref{gfactor}) that for elongated samples of fixed volume,
$G$ decreases roughly as the inverse sample length. Evaluating the
$G$-factor as explained in the preceding section gives $G = 0.03$
for Bi sample, and $G = 0.007$ for Au sample. Substitution of the
above estimates into Eq.~(\ref{main4}) gives the order of
magnitude of the noise spectrum power $10^{-15}\,{\rm V^2/Hz},$
and $10^{-16}\,{\rm V^2/Hz}$ for Bi and Au sample, respectively.
We see that in both cases the theoretical estimates are in a
reasonable agreement with the experimental data. A more detailed
comparison is a much more difficult task which would require
specification of the precise form of the function $\beta(z).$

\section{Conclusions}\label{conclude}

We have shown that in the whole measured frequency band, the power
spectrum of quantum fluctuations of the Coulomb field exhibits an
inverse frequency dependence [Cf. Eq.~(\ref{main4})]. This result
is obtained by evaluating the two-point correlation function of
the electromagnetic field within the long-range expansion. It was
proved, in particular, that the low-frequency asymptotic of the
Fourier transform of this function is dominated by its
disconnected part. Furthermore, we have argued that the derived
power spectrum possesses the other experimentally observed
properties of the $1/f$-noise as well. In particular, the exact
dependence of flicker noise on the sample geometry has been
established. That the power spectrum of flicker noise is
proportional to the sample volume only very approximately was
noticed long ago, but precise form of this dependence has not been
inferred. We have shown that for geometrically similar samples the
noise power spectrum is inversely proportional to the sample
volume, while for samples of the same volume dependence on the
sample geometry is described by the dimensionless $G$-factor given
by Eq.~(\ref{gfactor}).

Finally, according to the semi-quantitative estimates of
Sec.~\ref{estim}, the found noise level is in a reasonable
agreement with experimental data. Thus, quantum fluctuations of
the Coulomb field produced by elementary particles can be
considered as one of the underlying mechanisms of the observed
$1/f$-noise.

\acknowledgments{I thank Drs. G.~A.~Sardanashvili,
K.~V.~Stepanyantz (Moscow State University) for interesting
discussions, and especially P.~I.~Pronin for introducing me into
the problem of $1/f$-noise.}

\pagebreak

\centerline{\bf Figure captions}

Fig.1: Tree contribution to the right hand side of
Eq.~(\ref{fint}). Wavy lines represent photon propagators, solid
lines the scalar particle.

Fig.2: Tree contribution to the right hand side of
Eq.~(\ref{fintctp1}). Part (e) of the figure represents the
``transposition'' of diagrams (a)--(d) (see Sec.~\ref{discon}).

Fig.3: Voltage measurement in geometrically similar samples
($n=2$).

\pagebreak

\pagebreak

\pagebreak

\end{document}